\documentclass[twocolumn, showpacs, english, preprintnumbers, amsmath, amssymb, superscriptaddress, aps]{revtex4-1}
\usepackage{latexsym}
\usepackage{epsfig,psfrag}
\usepackage{dcolumn}
\usepackage{bm}
\usepackage{graphicx}
\usepackage{color}
\usepackage{esint}
\usepackage{babel}
\usepackage{amsfonts}
\usepackage{enumerate}
\usepackage{mathbbol}

\begin{document}
\title{Chiral interface states in graphene ${\bm p}$-${\bm n}$ junctions} 

\author{Laura Cohnitz}
\affiliation{Institut f\"ur Theoretische Physik, 
Heinrich-Heine-Universit\"at, D-40225 D\"usseldorf, Germany}

\author{Alessandro De Martino}
\affiliation{Department of Mathematics, City University London, 
London EC1V 0HB, United Kingdom}

\author{Wolfgang H\"ausler}
\affiliation{Institut f\"ur Physik, 
Universit\"at Augsburg, D-86135 Augsburg, Germany}
\affiliation{I. Institut f\"ur Theoretische Physik, 
Universit\"at Hamburg, D-20355 Hamburg, Germany}

\author{Reinhold Egger}
\affiliation{Institut f\"ur Theoretische Physik, 
Heinrich-Heine-Universit\"at, D-40225 D\"usseldorf, Germany}

\date{\today}

\begin{abstract}
We present a theoretical analysis of unidirectional interface states which 
form near $p$-$n$ junctions in a graphene monolayer subject to a homogeneous magnetic field.
The semiclassical limit of these states corresponds to trajectories propagating along the 
$p$-$n$ interface by a combined skipping-snaking motion. 
Studying the two-dimensional Dirac equation with a magnetic field
and an electrostatic potential step, we provide and discuss the exact
and essentially analytical solution of the quantum-mechanical eigenproblem for 
both a straight and a circularly shaped junction. The spectrum consists of localized Landau-like 
and unidirectional snaking-skipping interface states, where we always find 
at least one chiral interface state. For a straight junction and at energies
near the Dirac point, when increasing the potential step height, the group velocity of this 
state interpolates in an oscillatory manner between the classical drift velocity in a crossed 
electromagnetic field and the semiclassical value expected for a purely snaking motion. 
Away from the Dirac point, chiral interface states instead resemble the conventional skipping 
(edge-type) motion found also in the corresponding Schr\"odinger case. We also investigate the 
circular geometry, where chiral interface states are predicted to induce sizeable equilibrium ring currents. 
\end{abstract}

\maketitle

\section{Introduction}

The physics of two-dimensional (2D) graphene monolayers  
has been intensely studied over the past decade 
\cite{geim2004,geim2005,rmp1,goerbig,philip,eva,miransky}. 
A noteworthy recent development in this field is that the ballistic transport regime  (with mean free
paths beyond tens of $\mu$m) has become accessible, for instance, by using 
ultraclean suspended samples \cite{eva} or by encapsulating 
graphene layers in boron nitride crystals  \cite{boron}.  We here consider the electronic properties of 
graphene $p$-$n$ junctions in a perpendicular magnetic field $B$. This system has attracted considerable attention and many  interesting experimental 
transport studies have already appeared \cite{williams,huard,oezil,young,marcus,marcus1,
haug,Amet,ozyilmaz,Klimov,christian1,christian2,zhao,Matsuo,Kumada,tovari,
schen,lee2016,christian3}. In such setups, the gapless Dirac fermion spectrum of low-energy quasiparticles
in graphene \cite{rmp1,philip} allows for the  controlled electron or hole doping of parts of the sample
by backgate voltage changes. We mention in passing that one may also fabricate high-quality $p$-$n$ junctions in 
graphene by the controlled diffusion of metallic contacts \cite{Liu2016}.
Early experiments have reported a fractional quantization of the conductance across 
the $p$-$n$ junction  \cite{williams,huard,oezil,young,marcus,marcus1}. 
An explanation for this phenomenon is possible by taking into account chiral interface states 
propagating along the junction. 
For sufficiently disordered samples, their existence allows for a simple physical picture of the 
observed conductance quantization \cite{levitov,brouwer}.  
Over the past few years, experiments have approached the ballistic regime
\cite{haug,Amet,ozyilmaz,Klimov,christian1,christian2,zhao,Matsuo,Kumada,tovari,
schen,lee2016,christian3}, thereby realizing gate-controlled electron waveguides.

Motivated by the above developments, we here theoretically study chiral interface states for 
ballistic bipolar junctions of 2D Dirac fermions in a perpendicular homogeneous $B$ field.  
Such states have been analyzed on the semiclassical level in 
Refs.~\cite{carmier1,carmier2,Patel}.  The corresponding trajectories involve skipping orbits 
 combined with snake-type motions along the interface.
This can be rationalized by noting that (i) in electron vs hole doped regions, 
cyclotron orbits have different orientation sense, and that (ii) Klein tunneling allows for a finite 
probability $P(\theta)$ of an impinging particle to cross the $p$-$n$ 
junction, where $P$ depends on the incidence angle $\theta$ \cite{Patel}. 
With probability $1-P$, the particle is thus reflected back into the same region,  
resulting in the skipping motion of a conventional edge state. With probability $P$, 
however, the particle enters the other side according to Snell's law of negative refraction \cite{rmp1}.  
The cyclotron orbit is now traversed in opposite direction, and one obtains a snaking motion along
the junction.  Semiclassical trajectories are in general composed  of stochastic sequences of 
these elementary skipping/snaking units \cite{carmier1,carmier2,Patel}. 
The only exception is the case of normal incidence ($\theta=0$), 
where a pure snake motion is possible since Klein tunneling becomes perfect.
Recent experiments have reported evidence for this limit by injecting quasiparticles into
the $p$-$n$ junction from edge states perpendicular to the interface \cite{christian1}.   
While some aspects of the quantum mechanical spectrum for a straight junction have been discussed
in Refs.~\cite{zhang,bruder},  an exact and basically analytical solution of the problem has not been 
given to the best of our knowledge. 
Below we report qualitative differences to the results of Ref.~\cite{zhang} 
and also address the circular geometry.
We note in passing that a related interface state is expected 
without $p$-$n$ junction for inhomogeneous magnetic fields containing a line 
separating $B>0$ and $B<0$ regions, where counterpropagating Landau orbits on 
different sides conspire to yield a snake state \cite{lambert,tarun,Sim,lee2011}.
However, this purely magnetic snake state is different 
and has not been observed experimentally so far.  
For other theoretical studies more distantly related to the present work, 
we refer the reader to Refs.~\cite{falko2006,portnoi2010,portnoi2012,portnoi2014,jiang2016}.

In the present work, we shall discuss chiral interface states for two different 
types of $p$-$n$ junctions, see Fig.~\ref{fig1}, namely for a 
straight and for a circularly symmetric junction. The latter case is closely 
related to recent experiments, where circular junctions have been 
created by direct gating \cite{tovari,christian3}, by scanning 
tunneling microscopy (STM) tips \cite{zhao}, or by local 
manipulation of defect charges in the substrate \cite{lee2016}.
Using the established STM resolution capabilities in both space and energy,
experiments could monitor the eigenstates of this system in full detail, 
cf.~also Ref.~\cite{zhang}. For a circular $p$-$n$ junction with $B=0$, such
STM results have already been reported \cite{zhao,lee2016}, but to observe  
the chiral interface states of interest here, one needs to consider finite $B$. 
By superconducting quantum interference device (SQUID) microscopy \cite{allen,kirtley}, local current 
densities can be detected as well.  This method may 
provide direct access to the equilibrium ring currents expected in the
circular geometry due to interface states.  We mention in passing that
circular geometries have also been studied for electrostatic potentials
in Refs.~\cite{bardarson2009,portnoi2011,portnoi2015}.

We emphasize that all predictions below can be tested with existing experimental setups.  
Apart from graphene,
our results may also apply to the Dirac fermion surface states of topological 
insulators, cf.~Refs.~\cite{zhang,bruder,joel,alfredo}, or to the chiral
metal discussed in Refs.~\cite{chalker1995,chalker2000}.
Let us also mention that in $p$-$n$-$p$ or $n$-$p$-$n$ devices, exotic non-Fermi-liquid states 
are possible when electron-electron interactions between counterpropagating chiral interface states
are taken into account \cite{hausler2,laura}.  However, for the $p$-$n$ setups below, 
interaction effects are expected to be weak and will thus not be included.
In view of the high sample qualities nowadays achieved  in 
graphene monolayers, we assume a clean system which is free of disorder.
 
The structure of this article is as follows. In Sec.~\ref{sec2}, we summarize 
the Dirac fermion description underlying our analysis, followed by a discussion 
of the straight $p$-$n$ junction geometry in Sec.~\ref{sec3}.  The corresponding 
Schr\"odinger version \cite{rosenstein} is briefly reviewed in App.~\ref{appa}.
In Sec.~\ref{sec4}, we turn to the solution of the circular setup, 
where details of our perturbative analysis can be found in App.~\ref{appb}. 
Finally, we offer some concluding remarks in Sec.~\ref{sec5}.
Throughout this paper, we use units with $\hbar=1$.

\section{Model}\label{sec2}

We use the standard 2D Dirac-Weyl Hamiltonian to describe
low-energy quasiparticles in a graphene monolayer \cite{rmp1}, 
\begin{equation} \label{genham}
H= v_F \sigma_x \left(p_x+\frac{e}{c}A_x\right)+ v_F\sigma_y
\left(p_y+\frac{e}{c}A_y\right) + V\sigma_0,
\end{equation}
where $v_F\simeq 10^6$~m$/$s is the Fermi velocity and
$p_{x,y}=-i\partial_{x,y}$. The Pauli 
matrices $\sigma_{x,y,z}$ (with identity $\sigma_0$) 
act in the sublattice space corresponding to the two-atom basis  of the 
honeycomb lattice. 
A constant perpendicular magnetic field $B=\partial_xA_y-\partial_y A_x$ with $B>0$
is encoded by the vector potential $(A_x,A_y)$.
With minor adjustments, Eq.~\eqref{genham} also describes 
graphene's quasiparticles in
strain-induced pseudo-magnetic fields, see Ref.~\cite{rmp1}. 
The scalar potential $V(x,y)$ in Eq.~\eqref{genham} comes from 
electrostatic gating, where spatial variations are expected to be smooth
on the scale of the lattice spacing, i.e., 
$V$ does not scatter quasiparticles between different ($K/K'$) valleys.
Since the magnetic Zeeman term (not specified above) is diagonal in spin space 
and can be absorbed by an overall energy shift \cite{rmp1}, 
we keep both spin and valley indices implicit. 

\begin{figure}
\centering
\includegraphics[width=4cm]{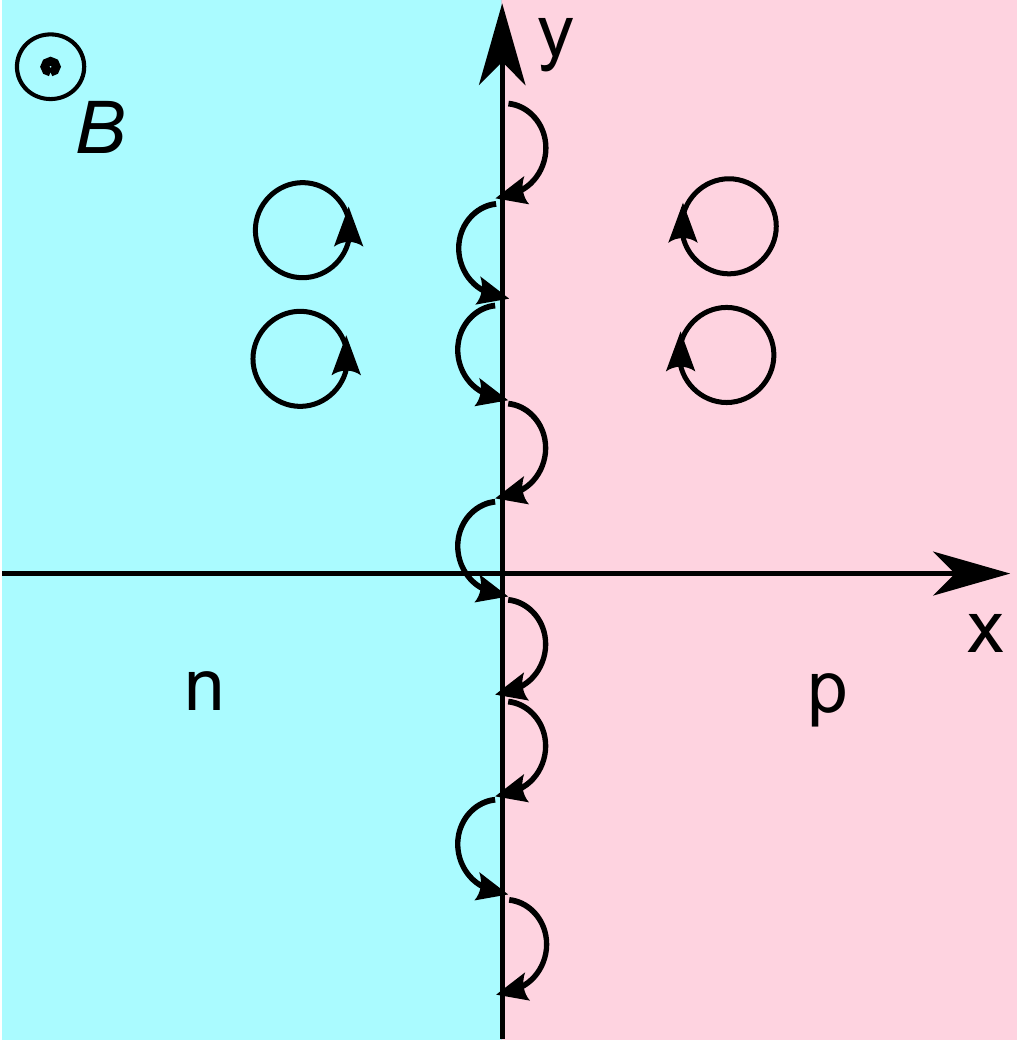} \hfill \includegraphics[width=4cm]{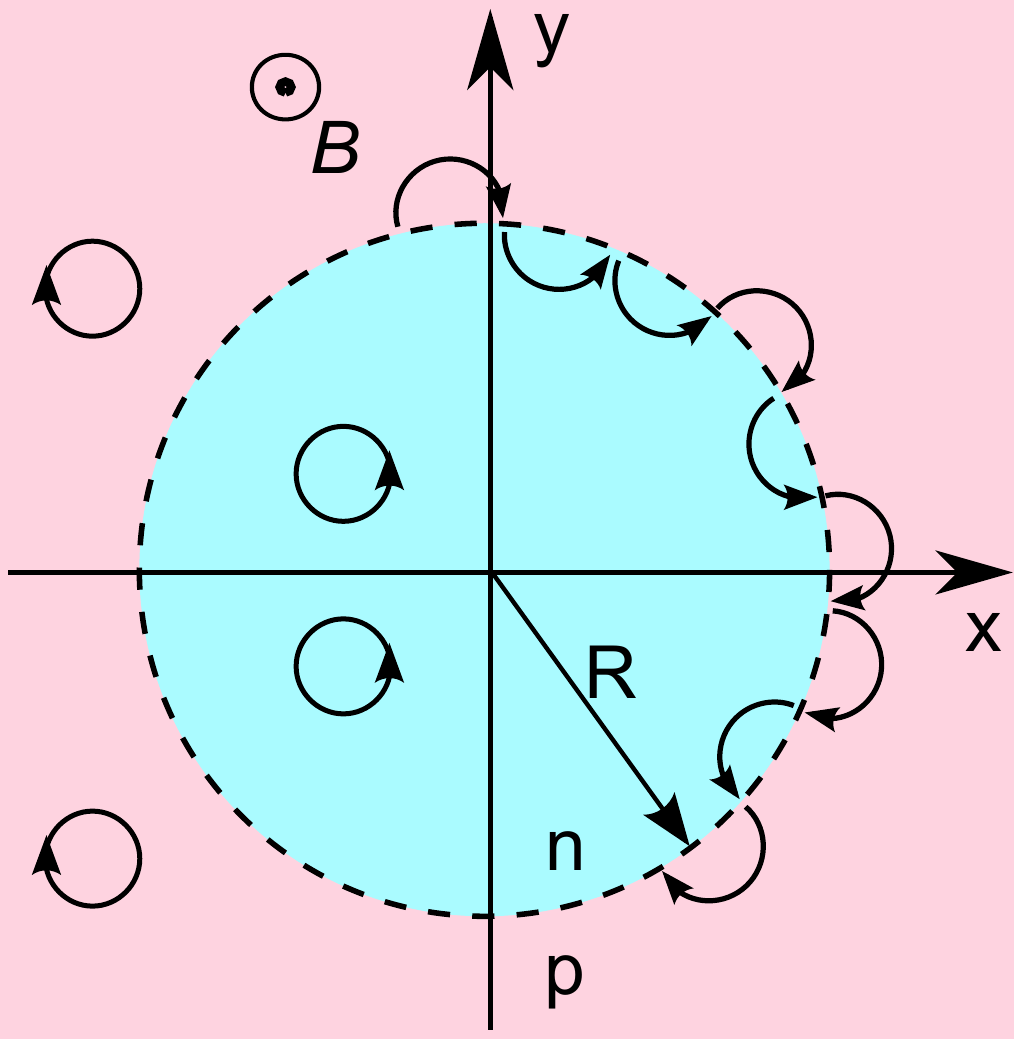}
\caption{\label{fig1} Sketch of two types of 
graphene $p$-$n$ junctions in a constant $B$ field, including 
examples for semiclassical cyclotron and/or skipping-snaking interface trajectories. 
Blue (red)  stands  for $n$-doped ($p$-doped) regions with 
constant potential $V=-V_0$ ($V=+V_0)$. 
Left panel: Straight junction, see Eq.~\eqref{straightpn}.  
Right panel: Circular junction, see Eq.~\eqref{circularpn}.}
\end{figure}

Within the above approximations, we are thus left with a single Dirac fermion 
species described by the Hamiltonian in Eq.~\eqref{genham}.
Below we specify all lengths (energies) in units of the 
magnetic length (energy) scale $l_B$ ($E_B$) with
\begin{equation}\label{units}
l_B= \sqrt{c/eB} ,\quad E_B=\sqrt{2} v_F/l_B.
\end{equation}
Moreover, we measure wave numbers $k$ in units of $l_B^{-1}$.
For a typical field of $B=4$~T, this gives $l_B\simeq 13$~nm and $E_B\simeq
62$~meV.  The relativistic Landau level energies for $V=0$ are then
given by $E^{(0)}_n={\rm sgn}(n) \sqrt{|n|}$ with integer $n$ \cite{rmp1}.  
We shall analyze the spinor eigenstates of $H$ in Eq.~\eqref{genham} for 
an infinite 2D graphene sheet with the two model potentials illustrated in Fig.~\ref{fig1}.

First, we will study a straight $p$-$n$ junction along the $y$-axis defined by the
antisymmetric potential step 
\begin{equation}\label{straightpn}
V(x) = V_0~ {\rm sgn}(x),
\end{equation}
with $V_0>0$.  In practice, such a potential step is created through the application of suitable gate voltages on both sides of the junction. 
The quantitative form of the electrostatically created potential can be 
estimated by solving Poisson's equation, and one finds that 
the length scale over which the electrostatic potential changes
 from $-V_0$ to $+V_0$ is of the order of the graphene-gate distance but
never falls below $l_B$ \cite{levitov}. 
We here consider the sharp step in Eq.~\eqref{straightpn}, which captures the essential physics and is easier to analyze \cite{zhang,bruder}.    

As second example, we will consider a circular $p$-$n$ junction of radius $R$ around the origin, 
\begin{equation}\label{circularpn}
V(x,y) = V_0~ {\rm sgn}(r-R),\quad r=\sqrt{x^2+y^2}.
\end{equation}
For numerical studies of related setups, see Refs.~\cite{zarenia,szafran}.
As depicted in Fig.~\ref{fig1}, we then expect to find ring-like interface states. 
Due to their unidirectional character, neither interference nor commensurability effects
are anticipated from their existence. However, a persistent equilibrium ring current should appear,
see Sec.~\ref{sec4} below.  
 
\section{Straight junction}\label{sec3}

Let us start with the case of a straight $p$-$n$ junction
along the $y$-axis, with $V(x)$ in Eq.~\eqref{straightpn}. 
In order to preserve translational invariance along the junction axis, 
we choose the Landau gauge for the vector potential, $(A_x,A_y)=(0,Bx)$. 
With the conserved wavenumber $k=k_y$, spinor  eigenstates take the form 
\begin{equation}\label{spinorsolution}
\Psi_k(x,y)= e^{iky} \Phi_k(x),\quad 
\Phi_k(x) = \left( \begin{array}{c} u_k (x) \\ i v_k (x) \end{array}\right),
\end{equation}
and Eq.~\eqref{genham} reduces to a 1D problem. 
Using the units \eqref{units}, the spinor components in 
Eq.~\eqref{spinorsolution} then satisfy 
\begin{equation}\label{1deq}
av =(E-V)u, \quad a^\dagger u=(E-V) v,
\end{equation} 
where the ladder operators $a=\partial_q+q/2$ and $a^\dagger= -\partial_q+q/2$
are defined in terms of a shifted 1D coordinate 
$q=\sqrt{2}(x+k)$.  These definitions imply the canonical commutator $[a,a^\dagger]=1$.

For a region of constant potential, $V(x)=V$, by eliminating $u$ from 
Eq.~\eqref{1deq}, we obtain
\begin{equation}\label{weber}
 \left[a^\dagger a- (E-V)^2\right] v =0,
\end{equation}
which is equivalent to Weber's equation \cite{abramowitz}. 
Using the recurrence relations of parabolic cylinder functions 
$D_p(q)$ \cite{abramowitz},
\begin{equation}\label{recurrence} 
a D_p(q)=p D_{p-1}(q),\quad a^\dagger D_p(q)=D_{p+1}(q),
\end{equation}
one directly verifies that Eq.~\eqref{weber} will be solved by 
$v=D_p(q)$ with the index $p=(E-V)^2$. 
Taking into account Eq.~\eqref{1deq}, the  spinor in
Eq.~\eqref{spinorsolution} thus follows (not normalized) as
\begin{equation} \label{phi1}
  \Phi_{k,V}^{(1)}(x) = \left(\begin{array}{c} (E-V)D_{(E-V)^2-1}(\sqrt2(x+k)) \\ 
iD_{(E-V)^2}(\sqrt2(x+k)) \end{array}\right).
\end{equation}
Noting that $q\to -q$ implies $a\to -a$ and $a^\dagger\to -a^\dagger$, i.e.,
$a^\dagger a$ remains invariant, we  observe that $D_p(-q)$ also solves Eq.~\eqref{weber}.
This yields a second spinor solution, 
\begin{equation}\label{phi2}
\Phi^{(2)}_{k,V}(x)=\left(\begin{array}{c}
 -(E-V)D_{(E-V)^2-1}(-\sqrt2(x+k)) \\ 
iD_{(E-V)^2}(-\sqrt2(x+k)) \end{array}\right).
\end{equation}
For the potential \eqref{straightpn}, using asymptotic properties of the $D_p(q)$ functions,
normalizable eigenstates must then be of the general form 
\begin{equation}\label{ansatz}
\Phi_k(x) = \left\{ \begin{array}{cc} c_< \Phi_{k,-V_0}^{(2)}(x), & x<0,\\  
c_> \Phi_{k,+V_0}^{(1)}(x) , & x>0,\end{array}\right.
\end{equation}
with $k$-dependent complex coefficients $c_{</>}$.  These coefficients are next determined by 
imposing a matching condition at the interface together with overall state normalization.

Continuity of the spinor at $x=0$ implies the energy quantization condition   
\begin{equation}\label{matching}
\Delta_k(E)= {\rm det}\left[\Phi_{k,-V_0}^{(2)}(0), \Phi_{k,+V_0}^{(1)}(0)\right]= 0.
\end{equation}
The solutions $E=E_{n,k}$ to Eq.~\eqref{matching} yield the spectrum, where the integer 
band index $n$ coincides with the Landau level index when $V_0=0$.
The spectrum obeys the symmetry relation
\begin{equation}\label{symmetryrel} 
E_{-n,-k}=-E_{n,k},
\end{equation}
which follows from $\Delta_k(E)=\Delta_{-k}(-E)$, cf.~Eqs.~\eqref{phi1}--\eqref{matching}.  This means that 
for each eigenstate with energy $E$, there is a mirror state with energy $-E$ and
opposite wavenumber.   For given $E_{n,k}$, with the corresponding eigenvector $(c_<,c_>)$ of 
the $2\times 2$ matrix in Eq.~\eqref{matching},   
the eigenstate $\Psi_{n,k}(x,y)$ follows from Eqs.~\eqref{spinorsolution} and \eqref{ansatz}
with subsequent normalization.
The probability density, $\rho_{n,k}(x)$, which is normalized to unity, and
the $y$-component of the particle current density, $J_{n,k}(x)$, 
associated to this eigenstate are given by \cite{rmp1}
\begin{equation} \label{currentdens}
\rho_{n,k}(x)=\Psi_{n,k}^\dagger \sigma_0 \Psi^{}_{n,k}, \quad J_{n,k}(x)= v_F \Psi_{n,k}^\dagger \sigma_y \Psi^{}_{n,k} .
\end{equation}

\begin{figure}
\centering
\includegraphics[width=9.1cm]{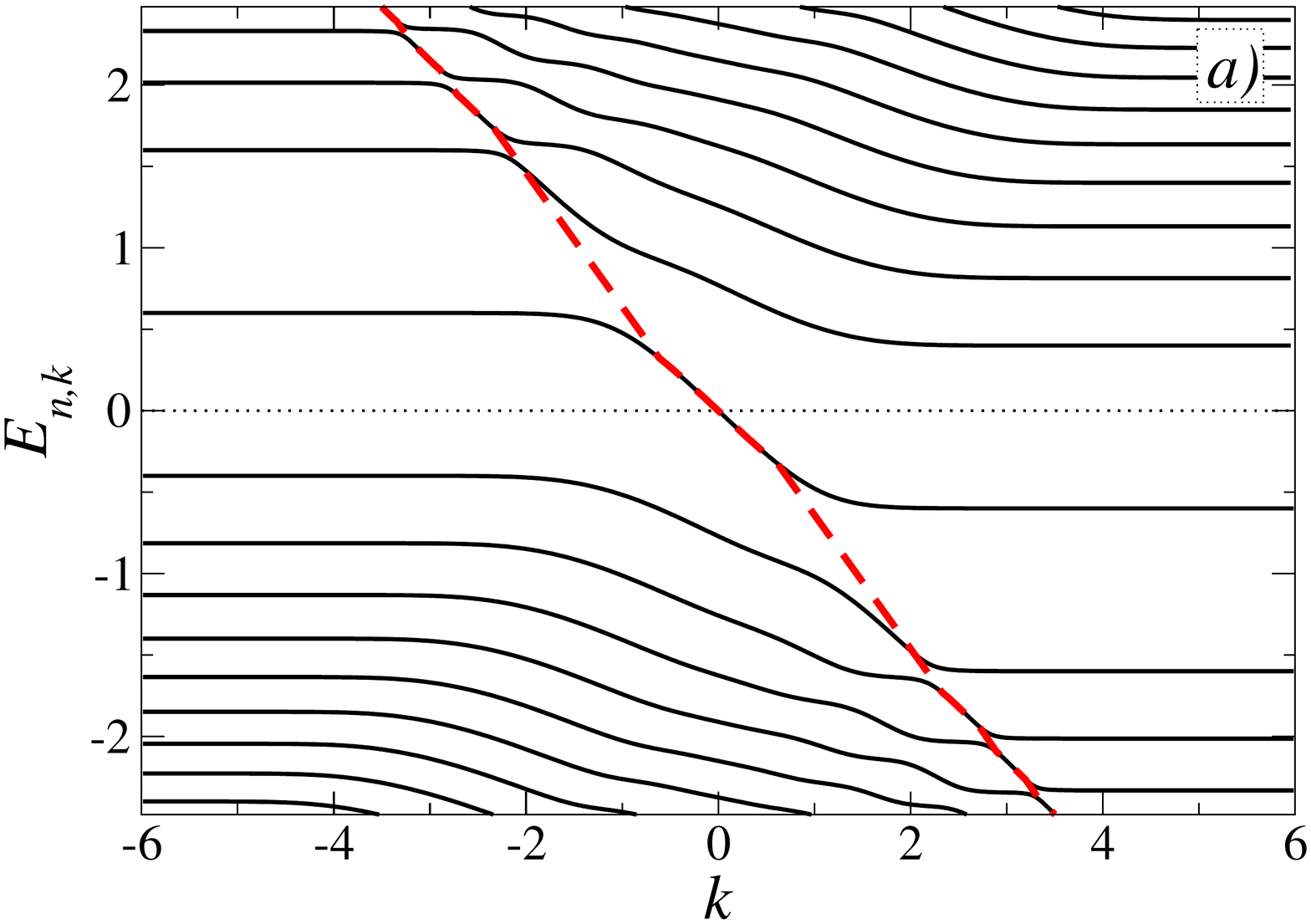}
\includegraphics[width=9.1cm]{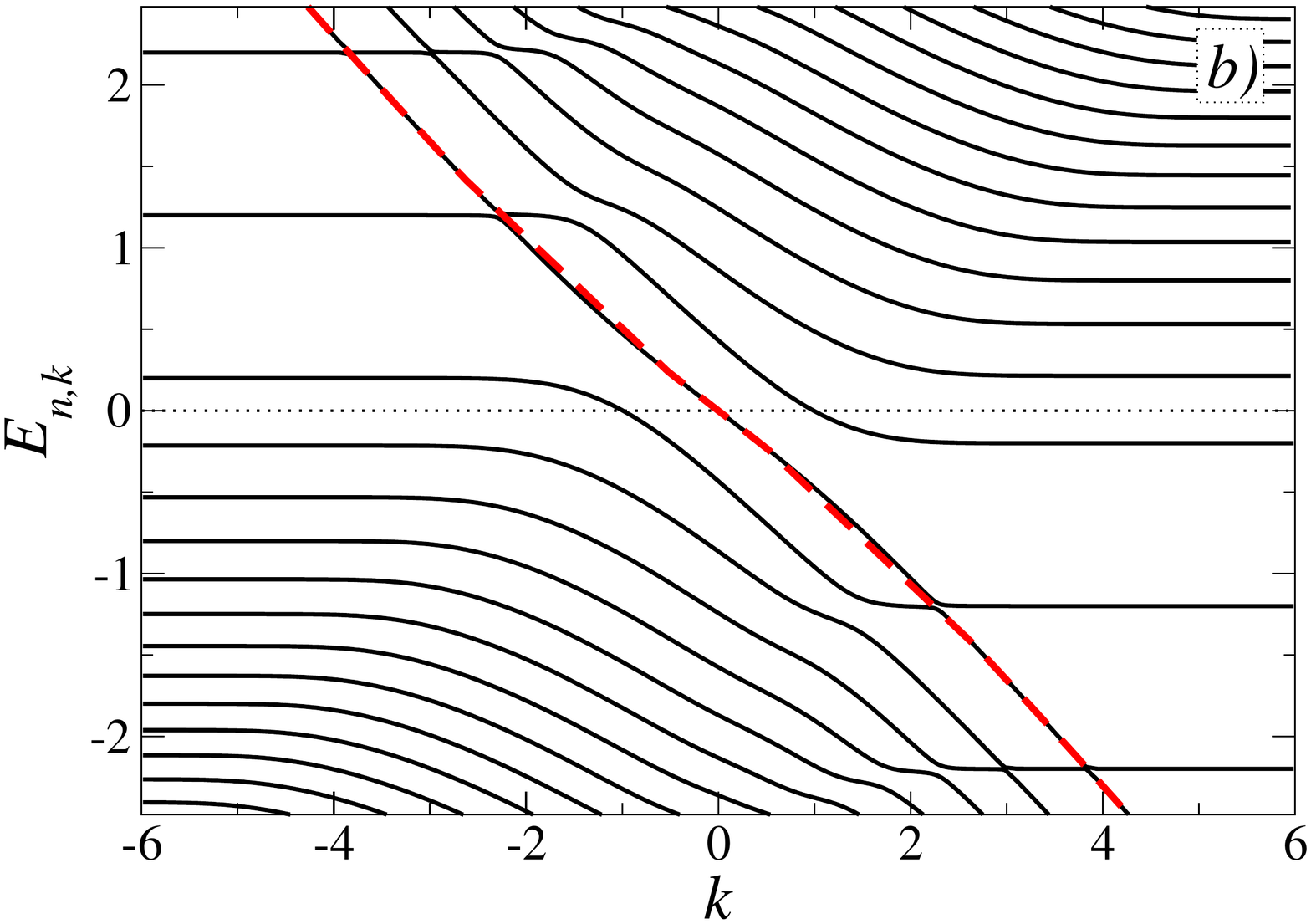}
\includegraphics[width=9.1cm]{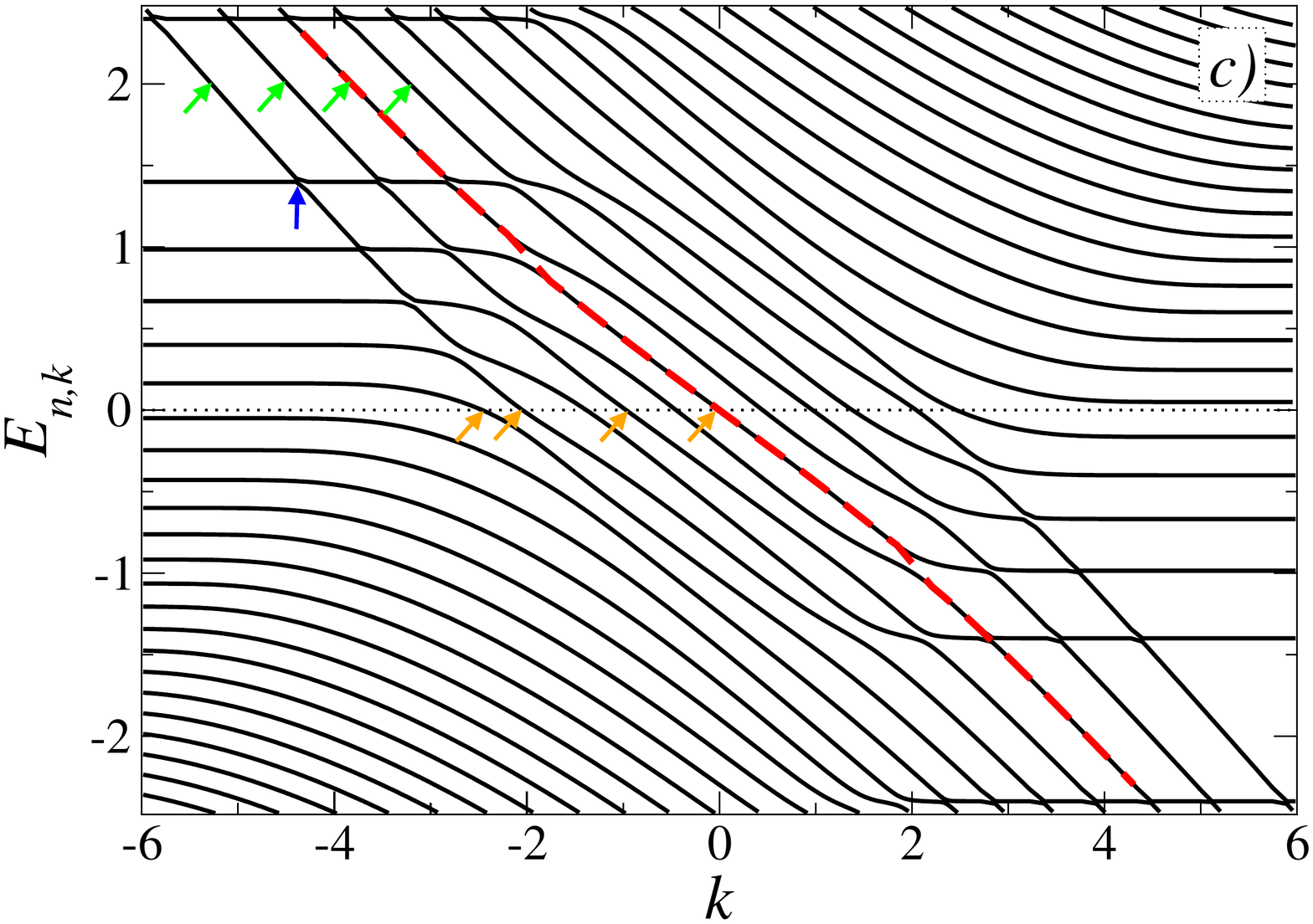}
\vspace*{-1cm}
\caption{\label{fig2} Spectrum $E_{n,k}$ vs $k$ obtained from Eq.~\eqref{matching} for 
a straight $p$-$n$ junction, with units in Eq.~\eqref{units}: (a) $V_0=0.6$, (b)  $V_0=1.2$,
(c) $V_0=2.4$. Red dashed curves are guides to the eye only and illustrate the 
central chiral interface state passing through $k=E=0$. Density profiles for the four 
$E=2$ ($E=0$) states labelled by green (orange) arrows are shown in Fig.~\ref{fig3} (in Fig.~\ref{fig4}).
The blue vertical arrow shows the avoided crossing studied in Fig.~\ref{fig5}.   
}
\end{figure}

The matching condition \eqref{matching} allows for analytical progress in certain limits, and 
in the general case  can be solved by numerical root finding (bracketing and bisection) techniques. 
For $V_0=0$, the 
solution of Eq.~\eqref{matching} reproduces the celebrated $k$-independent 
relativistic Landau level energies $E^{(0)}_{n,k}={\rm sgn}(n)\sqrt{|n|}$, where cyclotron orbits are centered around $\bar x=-k$. 

The exact spectrum for finite $V_0$, obtained numerically, is shown for three 
different values of $V_0$ in Fig.~\ref{fig2}.
Clearly, all energy bands $E_{n,k}$ contain flat parts for sufficiently large $|k|$.  
These parts correspond to Landau-like states centered far from, and thus
unaffected by, the interface, with the Landau energy shifted by $V_0$ 
(resp. $-V_0$) for $k<0$ (resp. $k>0$),  and corresponding cyclotron orbits 
centered at $\bar x>0$ (resp. $\bar x<0$).
Their dispersion follows by asymptotic expansion of Eq.~\eqref{matching} for $|k|\to \infty$.
By taking this limit at fixed energy, we obtain
\begin{equation}\label{asymptotic}
E_{n,|k|\to \infty}\simeq - {\rm sgn}(k) V_0 + {\rm sgn}(n)\sqrt{|n|},
\end{equation}
up to exponentially small $k$-dependent corrections.
 
The energy $E_{n,k}$ interpolates continuously between 
the two limits in Eq.~\eqref{asymptotic} when sweeping $k$
from $k\to -\infty$ to $k\to +\infty$. 
With increasing $V_0$, energy gaps between adjacent levels are thus progressively closed. 
The two largest gaps close simultaneously when 
the $n=0$ level for $k\to \mp \infty$ aligns with the $n=\pm 1$ level for $k\to \pm \infty$.
Using Eq.~\eqref{asymptotic}, this argument shows that for $V_0>1/2$, the 
entire spectrum becomes gapless.  

In addition to Landau-like states,  Fig.~\ref{fig2} shows that the bands
contain parts with negative slope which correspond 
to chiral interface states. The latter states are discussed in detail below.
They propagate with velocity 
\begin{equation}\label{velo}
\frac{v_{n,k}}{v_F}=  \sqrt{2} \frac{\partial E_{n,k}}{\partial k} = 
-\sqrt{2} \left(\frac{\partial_k \Delta_k(E)}{ \partial_E \Delta_k(E) } \right)_{E=E_{n,k}}
\end{equation}
along the $y$-axis, where the factor $\sqrt{2}$ is due to the units in Eq.~\eqref{units}.
The last expression, which specifies the velocity in terms of derivatives of $\Delta_k(E)$ 
in Eq.~\eqref{matching}, is convenient for numerical calculations of the velocity.

Before discussing the spectra in Fig.~\ref{fig2} in more detail, we note that 
exact analytical results can be obtained for the zero-energy solutions of Eq.~\eqref{matching} 
when choosing the specific potential strengths $V_0=\sqrt{N}$ (with $N=1,2,\ldots$). 
For positive integer index $p=N$ (including $N=0$), the parabolic cylinder functions appearing
in Eqs.~\eqref{phi1} and \eqref{phi2} reduce to conventional Hermite ($H_N$) polynomials by virtue
of the relation \cite{abramowitz}
\begin{equation}
D_N(q) = 2^{-N/2} e^{-q^2/4} H_N(q/\sqrt{2}).
\end{equation}
For $V_0=\sqrt{N}$, the matching equation $\Delta_k(E=0)=0$ then has the $2N-1$ solutions
$k= \{ 0, \pm k_1,\ldots,\pm k_{N-1} \}$,  where the $k_i$ are the $N-1$ positive zeroes of $H_N(k)$.  
Based on this argument, we conclude that for a potential strength within the bounds
\begin{equation} \label{bounds} 
\sqrt{N-1}<V_0\le \sqrt{N},
\end{equation}
there are $2N-1$ energy bands $E_{n,k}$ which cross $E=0$ at some $V_0$-dependent value of $k$. 
In particular, for $V_0\le 1$, there is a single band ($n=0$) which passes through
$E=0$ at $k=0$. 
We will show below that such bands correspond to the low-energy limit of chiral interface states.
This implies that for arbitrary $V_0$,  an odd number $2N-1\ge 1$ of these states exists, 
cf.~Eq.~\eqref{bounds}, on energy scales $|E|\alt 1$.

Let us now discuss the general case, starting from $V_0=0.6$ in Fig.~\ref{fig2}(a).  
Focussing on the level closest to zero energy, as $k$ increases, one interpolates 
between $n=0$ Landau states 
shifted by $\pm V_0$ on different sides of the junction, see Eq.~\eqref{asymptotic}, 
passing through chiral interface states with negative group velocity.  
These states are unidirectional, propagate only along the negative $y$-axis, and are 
centered near the $p$-$n$ junction at $x=0$.
 At the same time, interface states are also visible at higher energy scales, 
where they are formed  from $n\ne 0$ bands. 
For larger $V_0$, see Fig.~\ref{fig2}(b), narrower and narrower anticrossings separate 
regions of approximately linear dispersion originating from different bands. 
Therefore, as $k$ changes, one can identify a single chiral interface mode that evolves 
through a sequence of avoided crossings, where the band index $n$ changes along 
the way.  This mode is indicated by the red dashed line in Fig.~\ref{fig2}(b).
While the slope of the higher-energy part of the mode dispersion is approximately 
constant, we observe from Fig.~\ref{fig2} that the corresponding velocity $v_\infty>0$ 
(oriented along the negative $y$-axis) is slightly bigger than the 
velocity $v_s=-v_{n=0,k=0}$ observed near $E=0$. 
Of course, this interface mode is not a true eigenstate for all $k$.  
However, since the avoided crossings become very narrow, 
such a mode effectively represents eigenstates except for $k$-values near those anticrossings.
This scenario also applies to larger values of $V_0$, where 
more chiral interface modes can be identified, see Fig.~\ref{fig2}(c).

\begin{figure}
\centering
\includegraphics[width=8.65cm]{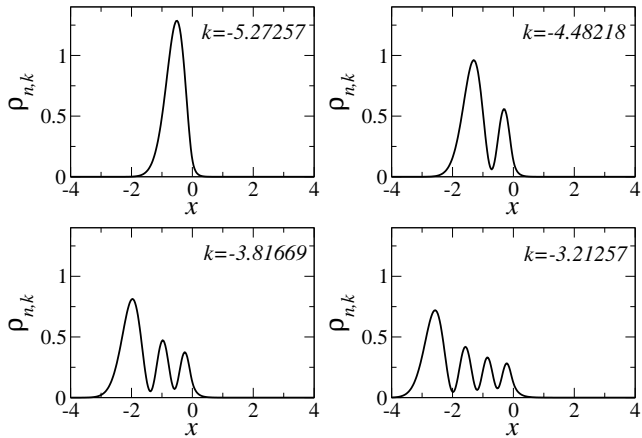}
\caption{\label{fig3} 
Spatial profile of the probability density $\rho_{n,k}(x)$ at $V_0=2.4$ 
 for four different eigenstates with energy $E_{n,k}=+2$
 and wavenumbers $k\simeq -5.27, -4.48, -3.82, -3.21$.
 These states are indicated by green arrows in Fig.~\ref{fig2}(c). }
\end{figure}

In Fig.~\ref{fig3}, we illustrate the probability density $\rho_{n,k}(x)$
defined in Eq.~\eqref{currentdens} for the four states with energy $E_{n,k}=+2$
indicated by green arrows in Fig.~\ref{fig2}(c). We observe that these states are 
located in close vicinity to the $p$-$n$ interface as compared to
the respective $V_0=0$ Landau states. For the shown $k$ values, the latter
states would be centered at $\bar x\simeq +5.27, +4.48, +3.82,$ and $+3.21$, i.e., 
further away and even on the other side of the interface.
Interestingly, the result for $k\simeq -5.27$ resembles
the (strongly displaced) probability density of a Landau state with $n=0$.  Indeed, 
Fig.~\ref{fig2}(c) confirms that this state continuously evolves to the $n=0$ shifted 
Landau level in Eq.~\eqref{asymptotic} when following its energy dispersion all the 
way to $k\to -\infty$. 
A similar observation holds true for the other $k$-values shown in Fig.~\ref{fig3}, 
which evolve to higher ($n=1,2,3$) Landau levels at $k\to -\infty$ through 
a series of $n+1$ avoided crossings.  
Note that the central chiral interface mode, which passes through $k=E=0$ and is 
highlighted as red dashed curve in Fig.~\ref{fig2}, corresponds to $n=2$ for the shown 
wavenumber $k\simeq -3.82$. 

Next we recall that for given $V_0$ within the bounds in
Eq.~\eqref{bounds}, $2N-1$ modes cross the $E=0$ line.  
From the shown numerical results, we infer that the dispersion relation for all these 
modes is linear at sufficiently low energy scales.   
For $V_0=1.2$, on top of the central chiral interface state which is always present, we observe that   
a pair of states reaches $E=0$ at finite wavenumbers $\pm k_1$. 
At low energies, those states are formed from $n=\pm 1$ bands,
where for $V_0>1$, the $n=-1$ ($n=+1$) shifted Landau energy moves above (below) zero energy 
for $k\to -\infty$ ($k\to +\infty$), cf.~Eq.~\eqref{asymptotic}.  
Furthermore, for $V_0=2.4$, Eq.~\eqref{bounds} predicts $2N-1=11$ zero-energy crossings
as confirmed by Fig.~\ref{fig2}(c). 
We conclude that for arbitrary $V_0$, there is always at least one chiral interface state present.
This conclusion is validated by the analytical observation that at $(E,k)=(0,0)$ 
the function $\Delta_k(E)$ vanishes but its partial derivatives are finite. As a consequence, 
the matching condition \eqref{matching} predicts a linear dispersion relation near $(k,E)=(0,0)$
for any value of $V_0$. 
The above results correct a finding of Ref.~\cite{zhang}, where interface 
states were argued to disappear for $V_0>1$ \cite{footnote}.

\begin{figure}
\centering
\includegraphics[width=8.65cm]{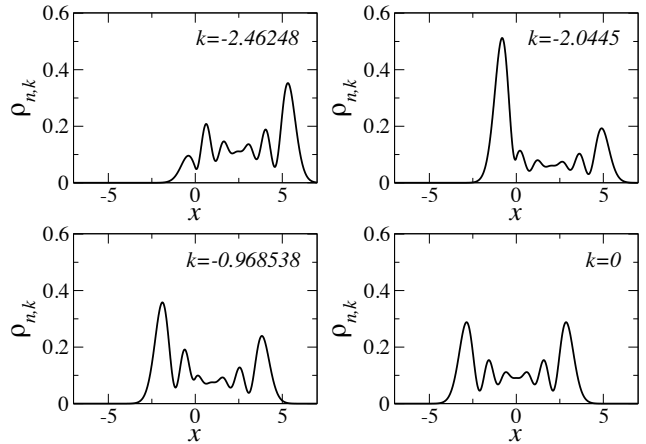}
\caption{\label{fig4}
Spatial profile of the probability density $\rho_{n,k}(x)$ at $V_0=2.4$  for the four 
zero-energy states with wavenumbers $k\simeq -2.46, -2.04, -0.97,$ and $k=0$, 
corresponding to  the orange arrows in Fig.~\ref{fig2}(c).} 
 \end{figure}

Interestingly, for the $E=0$ states illustrated in Fig.~\ref{fig4}, in particular when $|k|$ is small,
we observe that the probability density has finite weight on both sides of the interface,
in contrast to the $E=2$ states shown in Fig.~\ref{fig3}.  This feature is peculiar to the Dirac 
fermion nature of graphene quasiparticles, where low-energy states on the left/right side of the junction 
correspond to electrons and holes, respectively.  In fact, we 
explicitly show in App.~\ref{appa} that  the corresponding Schr\"odinger version of this problem 
does not contain such a state.  Figure \ref{fig4} demonstrates that the graphene $p$-$n$ interface state 
near $k=E=0$ has a spatially symmetric density profile, as expected for a pure snake motion.  
The asymmetric density profiles found at higher energies, see Fig.~\ref{fig3}, instead resemble the
edge-type interface states associated with skipping orbits in the semiclassical picture. The latter
type of chiral interface states are found also in the Schr\"odinger case, see App.~\ref{appa}.

\begin{figure}
\centering
\includegraphics[width=8.65cm]{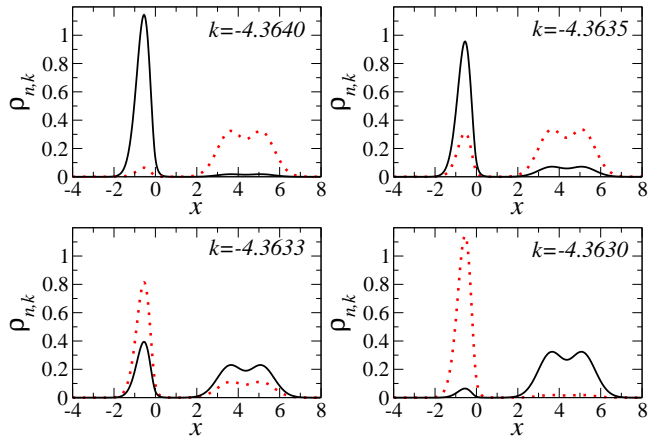}
\caption{\label{fig5}  
Spatial profile of the probability density $\rho_{n,k}(x)$ at $V_0=2.4$ for $n=-1$  (red dotted) and $n=0$ 
(black solid curves) states with wavenumbers $-4.364\le k\le-4.363$ near the avoided 
crossing indicated by the blue vertical arrow in Fig.~\ref{fig2}(c).  }
\end{figure}

We now address in more detail the metamorphosis between Landau and interface states when 
moving through an avoided crossing.  We illustrate this transition in Fig.~\ref{fig5}
by following the probability density through a specific avoided crossing.
While Landau-like states are centered near $\bar x= -k$, 
chiral interface states are located near the junction at $x=0$.
It is evident from Fig.~\ref{fig5} that the transmutation between 
Landau and chiral interface states happens over a very narrow region
of wavenumbers. The fact that the gap is so tiny can by rationalized by noting that both states 
are centered far from each other and therefore only come with a very small hybridization.

\begin{figure}
\centering
\includegraphics[width=9.5cm]{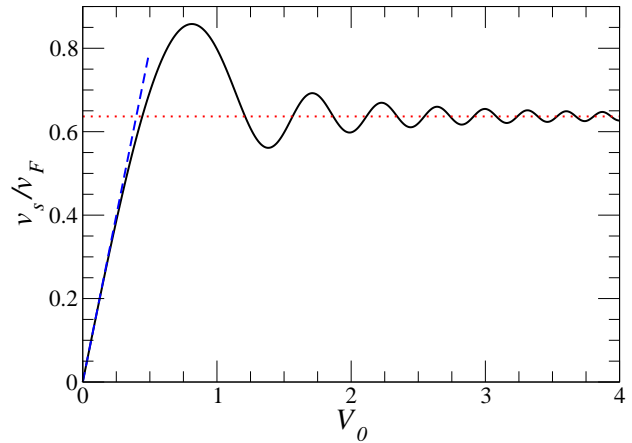}
\caption{\label{fig6} Velocity $v_s=-v_{n=0,k=0}$ of the central chiral interface state at low energy 
scales vs $p$-$n$ potential strength $V_0$.  
The solid black curve shows the analytical result in Eq.~\eqref{snakevelo}. 
The dotted red line denotes the large-$V_0$ limit, $v_s/v_F=2/\pi\simeq 0.63$. 
The dashed blue line gives the drift velocity $v_s/v_F=2\sqrt{2/\pi} V_0$ 
expected for $V_0\ll 1$. }
\end{figure}

Let us then turn to the velocity of the central chiral interface state passing through $k=E=0$. 
Our numerical results indicate different velocities at low and high energies, where
the dispersion relations for $|E|\agt 1$ and $|E| \alt 1$
have approximately constant velocities $v_\infty$ and $v_s$, respectively.  
A linear regression fit to the $E>1$ data in Fig.~\ref{fig2} gives the values
$v_\infty/v_F\simeq 0.95$ (0.92, 0.82) for $V_0=0.6$ (1.2, 2.4), 
consistent with the limiting behavior $v_\infty\to v_F$ expected for $|E|\gg V_0$ \cite{tarun}.
Our fitted values for $v_\infty$ are clearly larger than the respective velocities 
$v_s/v_F\simeq 0.76$ (0.65, 0.63) extracted from a linear regression fit near $E=0$. 
The latter numbers nicely match the analytical prediction
\begin{eqnarray} \label{snakevelo}
\frac{v_s}{v_F}&=& \frac{\sqrt{2\pi} \ 
2^{V_0^2}  V_0 \Gamma(1-V_0^2)}{1+V_0^2[\psi(1-V_0^2/2)-\psi(1/2-V_0^2/2)]} \\
\nonumber
&\times&\left( \frac{1}{\Gamma^2 (1/2-V_0^2/2)} - 
\frac{1}{\Gamma(-V_0^2/2)\Gamma(1-V_0^2/2)} \right),
\end{eqnarray}
with the Gamma function $\Gamma(z)$ and the Digamma function $\psi(z)=d\ln\Gamma/dz$ 
\cite{abramowitz}.
Equation \eqref{snakevelo} follows from Eq.~\eqref{velo} by expanding Eq.~\eqref{matching} around $k=E= 0$.

Three features of this result are particularly noteworthy. 
First, for $V_0\ll 1$,  Eq.~\eqref{snakevelo} predicts $v_s=(2/\sqrt{\pi}) V_0 l_B$, 
cf.~the dashed blue line in Fig.~\ref{fig6}.
This prediction is independent of the Fermi velocity $v_F$, but Fig.~\ref{fig6} 
shows that $v_s$ never exceeds $v_F$ for any value of $V_0$.
For not too strong magnetic fields, the quoted small-$V_0$
limit of Eq.~\eqref{snakevelo} is equivalent to the classical
drift velocity of a charged particle in crossed magnetic
$(B_z=B)$ and electric ($E_x$, with $|E_x|<B$) fields. The drift
velocity is then given by $v_y=cE_x/B_z$ along the negative
$y$-axis. Assuming that the potential drops across the junction
over a length of order $l_B$, the electric field at the interface
is $E_x\approx V_0/(el_B)$, and hence $v_y\approx V_0 l_B \approx
v_s$, see also Ref.~\cite{lukose}. Second, the velocity
oscillates as a function of $V_0/E_B\sim V_0\sqrt{B}$. By variation of 
backgate voltages and/or the magnetic field, $v_s$ can therefore be changed over a wide parameter region.  
The extrema in $v_s$ approximately occur for $V_0=\sqrt{N}$, where new interface states 
are generated and band mixing between $n=0$ and $\pm N$ bands becomes important. 
Third, for $V_0\gg 1$, Eq.~\eqref{snakevelo} predicts that the 
velocity approaches $v_s=(2/\pi)v_F$. 
Interestingly, the same value is semiclassically expected for quasiparticles impinging on the junction
 under normal incidence, which then propagate with velocity $v_F$ along 
a semicircular snake trajectory on alternating sides of the junction 
\cite{carmier1,carmier2,Patel}. The average velocity along the junction axis
 will thus be given by $v_s=(2/\pi)v_F$.

\begin{figure}
\centering
\includegraphics[width=9.75cm]{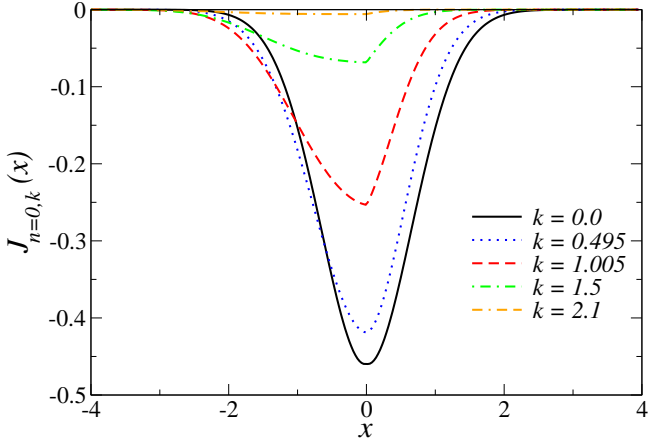}
\caption{\label{fig7} Current density profile $J_{n,k}(x)$, see Eq.~\eqref{currentdens}, 
for $n=0$ eigenstates of a straight graphene $p$-$n$ junction with $V_0=0.6$ and several wavenumbers $k$.  }
\end{figure}

Before turning to the circular geometry, we
finally address the particle current density, $J_{n,k}(x)$, along the $y$-axis, see 
Eq.~\eqref{currentdens}.  
Integrating over the transverse direction, 
the current  associated to a given eigenstate follows in the form (see also Ref.~\cite{tarun}) 
\begin{equation}\label{fullcurrent}
I_{n,k}= \int_{-\infty}^\infty dx J_{n,k}(x)= v_{n,k},
\end{equation}
with the velocity $v_{n,k}$ in Eq.~\eqref{velo}. The current is here measured
in units of $v_F/l_B$.
Note that $I_{n=0,k=0}$ is directly proportional to 
the velocity $v_s$ in Eq.~\eqref{snakevelo}, see Fig.~\ref{fig6}. 
For given chemical potential $\mu$, the total current is then given by 
$\sum_n \int dk \; I_{n,k} f(E_{n,k}-\mu)$, with the Fermi function $f(E)$.

The current density profile is  shown in Fig.~\ref{fig7} for $n=0$ states with several 
$k$-values, taking $V_0=0.6$, see also Fig.~\ref{fig2}(a) for the corresponding 
energy dispersion.
We first note that the particle current is always oriented along the negative $y$-axis, 
consistent with the negative current densities in Fig.~\ref{fig7}.  
For $k=0$, Fig.~\ref{fig7} shows that one has a symmetric current density 
profile, $J_{0,0}(x)=J_{0,0}(-x)$, and the maximum absolute value of the 
current $|I_{0,k}|$ is reached at this wavenumber. 
With increasing $|k|$, the overall current $|I_{0,k}|$ gradually decreases
and eventually becomes exponentially small in $k$, as expected for the Landau-like 
states formed at $|k|\gg V_0$.   During this process, the current density profile always 
retains a peak near the interface (i.e., at $x=0$) but
becomes more and more asymmetric. 
Remarkably, while the current density peak remains pinned to the interface, 
the probability density $\rho_{0,k}(x)$ (data not shown here) is 
centered further and further away from the interface as $|k|$ increases. 

\section{Circular junction}
\label{sec4}

We now address the case of a circularly symmetric potential with radius $R$, see Eq.~\eqref{circularpn}. 
Using polar coordinates with radial distance $r=\sqrt{x^2+y^2}$ and angle $\varphi$,  
 rotational symmetry is kept intact by taking  the symmetric gauge for the vector potential,
with vanishing radial part and azimuthal component $A_\varphi=Br/2$. 
Below it is convenient to use instead of $r$ the radial coordinate $\xi=r^2/2$, with 
$\xi_0=R^2/2$ for the position of the $p$-$n$ junction.
Spinor eigenstates of the Dirac equation with a circularly symmetric potential and the 
above vector potential are then labelled by the conserved half-integer angular momentum $j$.
With the integer band index $n$ labelling different solutions for given $j$,
and adopting the units in Eq.~\eqref{units}, their
explicit form is \cite{ademarti,ademarti2}
\begin{equation} \label{circstates}
\Psi_{n,j}(\xi,\varphi) = \frac{ \xi^{|j+\frac12|/2}e^{-\xi/2}}{\sqrt{2\pi}} 
\left(\begin{array}{c} e^{i(j-\frac12)\varphi}\phi_{n,j}(\xi)
 \\  i e^{i(j+\frac12)\varphi} \chi_{n,j}(\xi) \end{array}\right),
\end{equation}
where the radial functions $\phi(\xi)$ and $\chi(\xi)$ are normalized according  to 
\begin{equation}
\int_0^\infty d\xi \ \xi^{|j+\frac12|}e^{-\xi} \left(|\phi_{n,j}|^2+|\chi_{n,j}|^2\right)=1.
\end{equation}

In a region of constant potential $V$, the radial functions are expressed in terms of the confluent 
hypergeometric functions $\Phi(\alpha,\gamma;\xi)$ and $\Psi(\alpha,\gamma; \xi)$ \cite{abramowitz}.
With $V(\xi)=V_0 \ {\rm sgn}(\xi-\xi_0)$, the Heaviside step function $\Theta(x)$, 
complex coefficients $c_{</>}$, and keeping the index $n$ implicit, they are given with
$m=|j|+1/2$ as follows, see Ref.~\cite{ademarti,ademarti2}.
For $j>0$, we obtain
\begin{widetext}
\begin{equation}
\left(\begin{array}{c}\phi_j\\ \chi_j\end{array}\right)
= c_{<} \Theta(\xi_0-\xi) \left (\begin{array}{c} \frac{m}{\sqrt{\xi}} 
\Phi\left(m-(E+V_0)^2, m; \xi\right) \\
(E+V_0) \Phi\left(m-(E+V_0)^2, 1+m; \xi\right) \end{array}\right)  \label{circeigen}
+ c_{>}\Theta(\xi-\xi_0)  
\left (\begin{array}{c} \frac{E-V_0}{\sqrt{\xi}} \Psi\left(m-(E-V_0)^2, m; \xi\right) \\
\Psi\left(m-(E-V_0)^2,1+m; \xi\right) \end{array}\right) . 
\end{equation}
 For negative $j$, we instead find the eigenstates
\begin{eqnarray}
\left(\begin{array}{c}\phi_j\\ \chi_j\end{array}\right)&=& c_{<}\Theta(\xi_0-\xi) \left (
\begin{array}{c} \sqrt{\xi} (E+V_0) \Phi\left(1-(E+V_0)^2,1+ m; \xi\right) \\
-m\Phi\left(-(E+V_0)^2, m; \xi\right) \end{array}\right)\label{circeigenn} \\ \nonumber
&+&
 c_{>}\Theta(\xi-\xi_0)  \left (
\begin{array}{c}\sqrt{\xi} (E-V_0) \Psi\left(1-(E-V_0)^2,1+m; \xi\right) \\
\Psi\left(-(E-V_0)^2, m; \xi\right) \end{array}\right). 
\end{eqnarray}
Continuity of the spinor at $\xi=\xi_0$ then again gives an energy quantization condition
determining the spectrum, $E=E_{n,j}$.  
For $j>0$, we obtain this condition in the form
\begin{equation} \label{en1} 
(E-V_0) \left[ 1-\frac{d}{d\xi} \ln\Phi\left
(m-(E+V_0)^2,m;\xi\right)\Bigr|_{\xi=\xi_0}
\right]= (E+V_0) \left[ 1-\frac{d}{d\xi} \ln\Psi\left
(m-(E-V_0)^2,m;\xi\right)\Bigr|_{\xi=\xi_0}
\right], 
\end{equation}
while for $j<0$, it is given by
\begin{equation}\label{en2}
 (E-V_0)\frac{d}{d\xi} \ln
\Phi\left(-(E+V_0)^2,m;\xi\right)\Bigr|_{\xi=\xi_0}=(E+V_0)
\frac{d}{d\xi}\ln \Psi\left(-(E-V_0)^2,m;\xi\right)\Bigr|_{\xi=\xi_0}\:.
\end{equation}
\end{widetext}
We note in passing that for $E=0$ and integer values of $V^2_0$,
analytical solutions are possible again, in analogy to Sec.~\ref{sec3},
since the $\Phi$ and $\Psi$ functions can then be written 
in terms of Laguerre polynomials \cite{abramowitz}. 

\begin{figure}
\centering
\includegraphics[width=9.5cm]{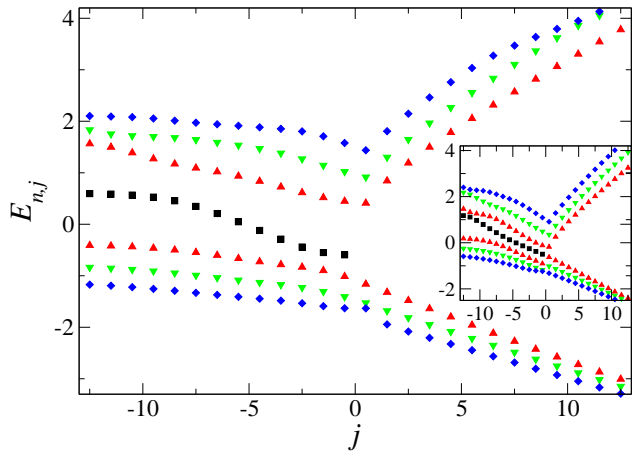}
\caption{\label{fig8}
Energy spectrum $E_{n,j}$ vs angular momentum $j$ for a circular $p$-$n$  
junction of radius $R=3.3$ with $V_0=0.6$ (main panel) and $V_0=1.2$
(inset), using the units in Eq.~\eqref{units}.
Different colors and symbols correspond to different values of $n_r=|n|-(j+1/2)\Theta(j)$:
$n_r=0$ (black squares), $n_r=1$ (red upward-pointing triangles) , 
$n_r=2$ (green downward-pointing triangles), $n_r=3$ (blue diamonds).}
\end{figure}

The spectrum, $E=E_{n,j}$, can now be obtained by numerical root finding methods and
is shown in Fig.~\ref{fig8} for $R=3.3$ and two values of $V_0$.  For
 $V_0=0$, the spectrum follows analytically as 
 \begin{equation}\label{circ0}
 E_{n,j}^{(0)}={\rm sgn}(n)\sqrt{|n|},\quad  |n|=n_r+(j+1/2)\Theta(j),
\end{equation}
with the radial quantum number $n_r\ge 0$. The zero-energy Landau level 
with $n=0$ is spanned by states with $n_r=0$ and $j<0$. Generally, $E^{(0)}_{n,j<0}$  
will be $j$-independent, while  $E^{(0)}_{n,j>0}\sim\sqrt{j}$ when $n_r$ is held fixed. 

For finite $V_0$, Fig.~\ref{fig8} shows that the spectrum does not change qualitatively for $j>0$, 
up to an overall energy shift  $V_0$ due to the positive potential
contribution for $r>R$. Indeed, we find $E_{n,j\gg 1} \approx V_0+E_{n,j}^{(0)}$.
On the other hand, the energies $E_{n,j<0}$ differ more substantially from Eq.~\eqref{circ0}. 
In particular, the $n=0$ level now acquires an angular momentum dependence, 
cf.~the black squares in Fig.~\ref{fig8}. 
We find a maximum (negative) slope of the $j$-dispersion at $j= -11/2$ for $V_0=0.6$,
see Fig.~\ref{fig8}. For the larger value $V_0=1.2$, cf.~inset of Fig.~\ref{fig8}, the corresponding 
value is at $j= -17/2$.  As we discuss below, this maximum slope is directly relevant for the 
experimentally observable ring current flowing around the $p$-$n$ interface.

\begin{figure}
\centering
\includegraphics[width=10cm]{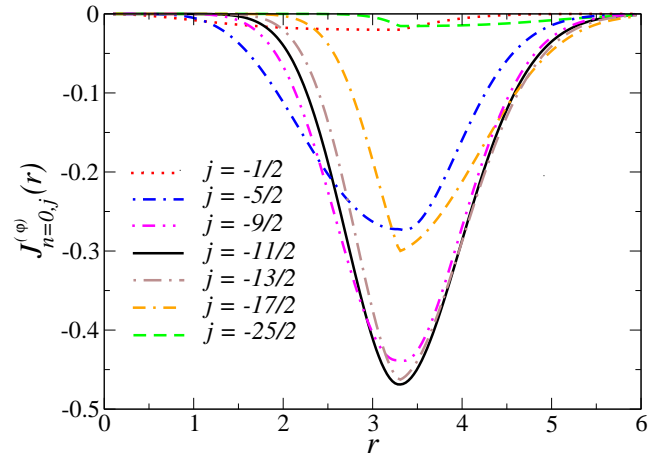}
\caption{\label{fig9} 
Current density in azimuthal direction, $J^{(\varphi)}_{n,j}(r)$, vs radial distance $r/l_B$ 
for a circular $p$-$n$ junction, see Eq.~\eqref{currazi}. The shown results 
are for $n=0$ and several $j<0$, with $V_0=0.6$ and $R=3.3$ as in the main panel of Fig.~\ref{fig8}. }
\end{figure}

The circulating current carried by a specific eigenstate $\Psi_{n,j}(r,\varphi)$ is defined by 
\begin{equation}\label{fullcurrent2}
I_{n,j}=\int_0^\infty dr J^{(\varphi)}_{n,j}(r),
\end{equation} 
with the current density  
\begin{equation}\label{currazi}
J^{(\varphi)}_{n,j}(r) = v_F \Psi_{n,j}^\dagger\left(\begin{array}{cc}
0 & -ie^{-i\varphi} \\ ie^{i\varphi} & 0 \end{array}\right) \Psi^{}_{n,j}
\end{equation}
running along the azimuthal direction.  The current density in the radial direction vanishes identically.
Equation \eqref{currazi} depends only on the radial variable $r$ and 
is shown in Fig.~\ref{fig9} for $n=0$ states with $j<0$. 

In analogy to Eq.~\eqref{fullcurrent}, $I_{n,j}$ can be written as angular 
momentum derivative of the dispersion relation, see also Ref.~\cite{halperin},
\begin{equation}\label{jderivative}
I_{n,j}=\frac{\sqrt2}{2\pi } \partial_j E_{n,j},
\end{equation} 
where the derivative is taken at fixed $n_r$. 
We note that the current is measured in units of $v_F/l_B$, where
the factor of $\sqrt{2}$ in Eq.~\eqref{jderivative} 
is again due to the units in Eq.~\eqref{units}.
The remarkable relation \eqref{jderivative} shows 
that the equilibrium ring current $I_{n,j}$ carried by an arbitrary 
eigenstate $\Psi_{n,j}$ is linked to the angular momentum dependence of the respective eigenenergy.   
For the $n=0$ state with $j=j_0<0$ where the steepest slope $\partial_j E_{0,j}$ is realized, 
the magnitude of the circulating current will thus be maximal.
This suggests that equilibrium ring currents due to chiral interface states are most pronounced 
when the Fermi level is aligned with $E_{0,j_0}$.

\begin{figure}
\centering
\includegraphics[width=9cm]{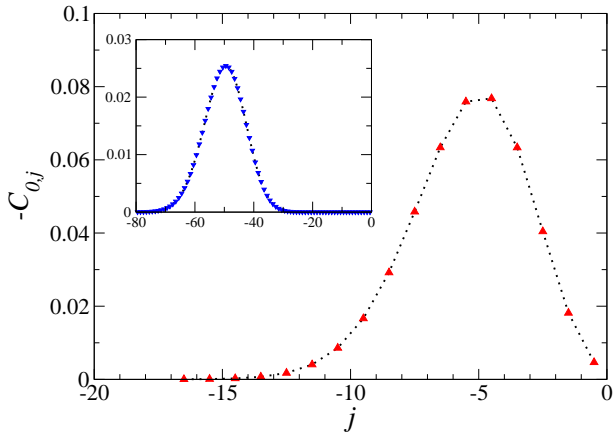}
\caption{\label{fig10}
Dimensionless coefficients $-C_{n=0,j<0}$ vs $j$ determining the ring currents 
in Eq.~\eqref{zerocurrent} for a circular $p$-$n$ junction with $R=3.3$ (main panel) and 
$R=10$ (inset). The shown results follow from first-order perturbation theory in $V_0$, see 
Eq.~\eqref{cpert} in App.~\ref{appb}. Dotted curves are guides to the eye only.  
In the inset, one clearly sees that the maximum of $|I_{0,j}|$ in Eq.~\eqref{zerocurrent} 
is reached for half-integer $j=j_0\approx -R^2/2 =-50$.}
\end{figure}

Noting that $I^{(0)}_{0,j<0}=0$, see Eq.~\eqref{jderivative} and Ref.~\cite{laura}, 
the current for small $V_0$ is given by
\begin{equation}\label{zerocurrent}
I_{0,j<0}=\frac{v_F V_0}{l_B E_B} C_{0,j} (\xi_0)+ {\cal O} (V_0^2).
\end{equation}
Here we have restored energy ($E_B$) and length
($l_B$) units in order to highlight that the magnetic field strength $B$ enters
the ring current \eqref{zerocurrent} only via the dependence of the dimensionless coefficients 
$C_{0,j}$ on the magnetic flux $\xi_0=\frac12(R/l_B)^2$ (in units of the magnetic flux quantum) 
through the $n$-doped disk region in Fig.~\ref{fig1}.  These coefficients are negative and
can be obtained analytically from perturbation theory in $V_0$, see App.~\ref{appb} and Fig.~\ref{fig10}, or numerically by means of Eq.~\eqref{jderivative}. Their absolute value is peaked at $j=j_0\approx -\xi_0$ with 
\begin{equation}\label{peakvalue}
|C_{0,j_0}|\approx \frac{0.25}{\sqrt{2\xi_0}}= \frac{0.25 l_B}{R}.
\end{equation} 
It is instructive to compare the magnitude of the maximum current (\ref{zerocurrent}), reached
at $j=j_0$,  
 to the corresponding maximum value of a conventional mesoscopic 
persistent current, $I^{(pc)}=v_F/(2\pi R)$.  This quantum-mechanical current flows in equilibrium
through a clean ring of radius $R$ threaded by a magnetic flux and depends on the magnetic field strength in an oscillatory manner \cite{hund}. 
The magnetic moment induced by the persistent current has
been measured by means of SQUID techniques \cite{mailly}.
Using Eqs.~\eqref{zerocurrent} and \eqref{peakvalue}, we find
\begin{equation}
\frac{|I_{0,j_0}|}{I^{(pc)} } \approx \frac{\pi}{2} \frac{V_0}{E_B}.
\end{equation}
We conclude that ring currents due to chiral interface states, 
as well as the thereby generated magnetic moments, will be sizeable at not overly small $V_0/E_B$. 

More detailed information can be obtained by measuring the spatially resolved 
current density distribution of the $n=0$
level, for example by using the experimental techniques employed in Refs.~\cite{allen,kirtley}.
Such profiles are displayed in Fig.~\ref{fig9} for $R=3.3$ and $V_0=0.6$.
Different values of $j$ can be addressed in experiments by aligning 
the Fermi energy with $E_{0,j}$, see Fig.~\ref{fig8}.  
As a function of the radial variable $r$, we observe from Fig.~\ref{fig9} that 
the current density exhibits a clear maximum near $r=R$, which is caused by circulating
chiral interfacial currents.  Indeed,
for $V_0\lesssim 1$, we find that the eigenstates $\Psi_{0,j<0}$ 
have a similar maximum also in the probability density $\rho_{0,j}(r)$ 
near $r=R$ and for half-integer $j=j_0\approx -R^2/2$. For larger $V_0$, 
however, oscillations around $r=R$ rather than a pronounced maximum are observed
both in the probability density and in the current density.

In Fig.~\ref{fig7}, we have shown similar chiral interfacial currents  for a 
straight junction with  the same potential strength $V_0=0.6$ as in Fig.~\ref{fig9}. This analogy 
between the straight and the circular geometry also applies to the spatial asymmetry
of the observed current density profiles. 
Finally, the total current $I_{0,j}$ follows by integrating the respective curve in Fig.~\ref{fig9}, 
see Eq.~\eqref{fullcurrent2}.  Clearly, for the chosen value of $R$ in Fig.~\ref{fig9}, 
the current is biggest for $j=j_0=-11/2$. By measuring ring currents for different 
choices of the Fermi level, one may thus be able to assign angular momentum numbers $j$ to 
individual quantum states.

\section{Conclusions}\label{sec5}

In this paper, we have given the full solution of the spectral problem 
for graphene $p$-$n$ junctions in an orbital magnetic field, studying both a straight junction 
and a circular geometry and focussing on the unidirectional interface states.
 
For a straight junction with a potential step of height $2V_0$, we have shown that there is
always an odd number of interface modes propagating in the same direction. 
By comparing the solution of the Dirac equation to the one for the corresponding Schr\"odinger
equation, we see that the graphene case is distinguished by the presence of a special
snake-type mode.  In addition,  both problems may feature common edge-state like modes. 
For small $V_0$, we find just one chiral interface mode for the graphene setup. 
This mode propagates in the low-energy limit with a group velocity set 
by the drift velocity in crossed electric/magnetic fields. For larger $V_0$, the velocity 
depends in an oscillatory manner on $V_0$, but saturates
at the semiclassical value $(2/\pi) v_F$ expected for a pure snake motion.

For the circular junction, we find qualitatively related results.
Chiral interface states can be controlled by the potential height
$V_0$ and the radius $R$ in their dominant angular momentum. Particularly interesting is the
zeroth Landau level, where a detectable ring current, which also causes a magnetic moment, 
will be induced by the chiral interface mode. For not too small $V_0$, this current is predicted
to be comparable in magnitude to the maximum persistent current 
flowing in a quantum ring of the same diameter. 
Furthermore, we have shown that the corresponding current density
is localized near the interface of the circular $p$-$n$ junction.

To conclude, we hope that our predictions can soon be put to an experimental test.  
This should be possible in available devices by using STM techniques and/or SQUID microscopy.  

\acknowledgments
This work was supported by the network SPP 1459 of the Deutsche Forschungsgemeinschaft (Bonn).

\appendix
\section{Chiral interface states for Schr\"odinger fermions}\label{appa}

\begin{figure}
\centering
\includegraphics[width=9cm]{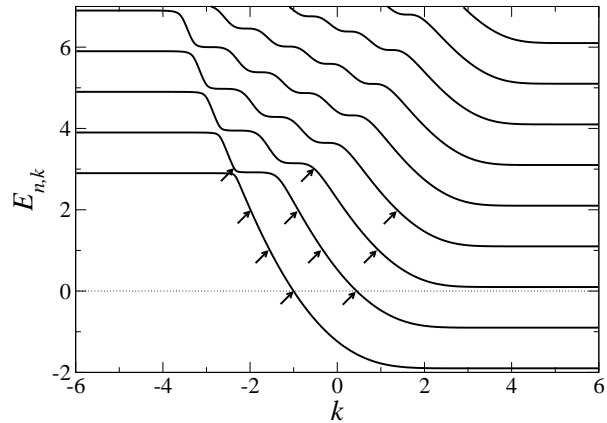}
\caption{\label{fig11}
Spectrum $E_{n,k}$ vs $k$ for a straight $p$-$n$ junction of Schr\"odinger fermions. 
Energies (wavenumbers) are in units of $\omega_c$ ($l_B^{-1}$), cf.~Eq.~\eqref{units}.
The shown results are for $V_0/\omega_c=2.4$ and follow from Eq.~\eqref{NRmatching}.
Density profiles are illustrated in Fig.~\ref{fig12} for the states with $E_{n,k}=0,1,2,3$ indicated by arrows.}
\end{figure}

Here we briefly discuss the chiral interface states in a straight $p$-$n$ junction
as in Sec.~\ref{sec3} but for Schr\"odinger fermions as in a conventional 2D electron gas, see Ref.~\cite{rosenstein}.
With the Landau gauge and the potential $V(x)$ in Eq. \eqref{straightpn}, eigenstates are as
in Eq.~(\ref{spinorsolution}),  $\Psi_k(x,y) = e^{iky} \psi_k(x)$, but with a scalar 1D wave function $\psi_k(x)$.
Using the same notation as in Sec.~\ref{sec3}, instead of Eq.~\eqref{1deq}, we now arrive at the 1D equation
\begin{equation}\label{1DSc}
\left[ a^\dagger a- \left( E -V(x) -\frac{1}{2} \right) \right] \psi_k(x)=0,
\end{equation}
where energies are measured in units of the cyclotron energy $\omega_c = eB/(mc)$ 
instead of $E_B$. For a region of constant potential $V(x)=V$, 
the two independent solutions of Eq.~\eqref{1DSc} are given by
\begin{equation}
\psi^{(1,2)}_{k,V}(x)=D_{p}(\pm\sqrt{2}(x+k)),\quad p=E-V-1/2.
\end{equation}
For a globally uniform potential $V$, normalizability implies
$p= n =0,1,2 \dots$, resulting in the standard (shifted by $V$) Landau level 
energies $E_{n,k}^{(0)}= n+1/2+V$.
For the potential in Eq.~\eqref{straightpn}, normalizable eigenstates take the general form,
cf.~Eq.~\eqref{ansatz},
\begin{equation}
\psi_k(x) =\left\{  \begin{array}{cc} c_< \psi^{(2)}_{k,-V_0}(x), & x<0, \\
c_> \psi^{(1)}_{k,+V_0}(x), & x>0. \label{NRwf}
\end{array} \right.
\end{equation}

\begin{figure}
\centering
\includegraphics[width=9cm]{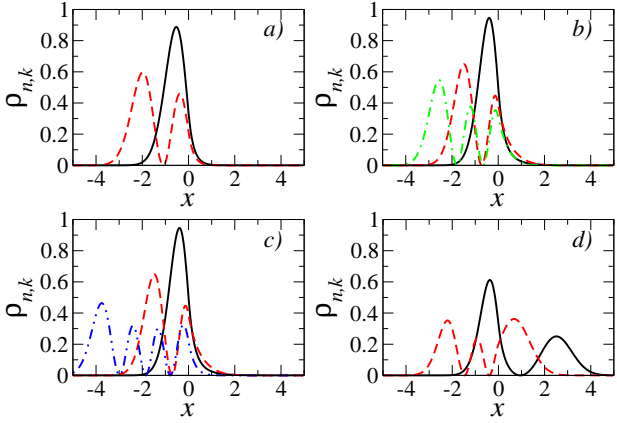}
\caption{\label{fig12}
Density profiles for a straight Schr\"odinger $p$-$n$ junction with $V_0/\omega_c=2.4$. 
The shown states are indicated by arrows in Fig.~\ref{fig11}, with $x$ in units of $l_B$.
 a) Energy $E_{n,k}=0$ with $k=-1.00205$ (black solid) and $k=0.445883$ (red dashed).
b) Energy $E_{n,k}=1$ with 
 $k=-1.54421$ (black solid), $k=-0.322044$ (red dashed),  and $k=0.953544$ (green dot-dashed).
c) Energy $E_{n,k}=2$ for
$k=-1.99022$ (black solid),  $k=-0.899269$ (red dashed), and  $k=1.37079$ (blue dot-dashed). 
d) Energy $E_{n,k}=3$ with $k=-2.35387$ (black solid) and $k=-0.519357$ (red dashed).
}
\end{figure}

The matching condition now involves the continuity of $\psi_k(x)$ and  $\psi'_k(x)$ at $x=0$.
By using recurrence relations for parabolic cylinder functions \cite{abramowitz}, one arrives at 
\begin{equation}
\Delta^{(S)}_k(E)=\det \left[ 
\begin{array}{cc}
\psi^{(2)}_{k,-V_0}(0) & -\psi^{(1)}_{k,V_0}(0)  \\
\psi^{(2)}_{k,-V_0-1}(0) & \psi^{(1)}_{k,V_0-1}(0)
\end{array}
\right] =0. \label{NRmatching}
\end{equation}
The solutions of Eq.~\eqref{NRmatching} determine the spectrum, $E=E_{n,k}$, where the 
band index $n$ again reduces to the Landau level index when $V_0=0$. 
The corresponding eigenfunction to energy $E_{n,k}$ is given by Eq.~\eqref{NRwf} with
$c_{</>} = {\cal N} D_{E_{n,k} \mp V_0-1/2} (\pm \sqrt{2}k)$
and overall normalization constant ${\cal N}$. 
The resulting spectrum is illustrated in Fig.~\ref{fig11}.
We observe that for finite $V_0$, Landau levels show
dispersion, where wide regions of approximately linear dispersion correspond
to chiral interface states. Moreover, we notice that avoided crossings appear again.

The probability density $\rho_{n,k}=\psi^*_{n,k}\psi^{}_{n,k}$ is illustrated in Fig.~\ref{fig12}
for several states with approximately linear dispersion relation, i.e., for chiral interface states.
We observe that these probability densities are always confined to
one side of the junction, with exponentially small weight on the other side. This is the behavior 
expected for edge-type states skipping along the junction as if it were a boundary.  
However, when the energy approaches a region of flat dispersion, e.g., near
an avoided crossing, the probability density exhibits finite weight on both sides of 
the interface, cf.~Fig.~\ref{fig12}(d), since here  Landau-type states coexist with chiral interface states. 

\section{Perturbation theory for circular geometry}\label{appb}

Here we discuss the results of perturbation theory in $V_0$ for the circular geometry
in Sec.~\ref{sec4}, where we obtain the coefficients $C_{0,j<0}$ in Eq.~\eqref{zerocurrent}
in closed analytical form.  For small $V_0\ll 1$, analytical progress for $I_{0,j}$ in Eq.~\eqref{fullcurrent2} 
is achieved by writing Eq.~\eqref{circularpn} as $V(r)=V_0+V^{\rm pert}(r)$ and 
treating $V^{\rm pert}(r)=-2V_0 \Theta(R-r)$ as small perturbation. 

Using the notation of Sec.~\ref{sec4}, cf.~Eq.~\eqref{circstates},  
and noting that the perturbation $V^{\rm pert}$ does not couple states with 
different angular momenta,
the $n=0$ Landau level states with $n_r=0$ and $j<0$
are given to lowest order in $V^{\rm pert}$ by
\begin{eqnarray}
&&\left(\begin{array}{c}\phi_{0,j}^{\rm pert}(\xi)\\
\chi_{0,j}^{\rm pert}(\xi)\end{array}\right)=
\frac{\sqrt{2}}{N_{0,j}}\left(\begin{array}{c} 0\\ 1 \end{array} \right)+
\sum_{n\ne 0}\frac{\left\langle n,j|V^{\rm pert}|0,j
\right\rangle} {-{\rm sgn}(n)\sqrt{|n|}N_{n,j}} 
 \nonumber \\ &&\qquad \times \label{psipert}
\left(\begin{array}{c}
\frac{-{\rm sgn}(n)\sqrt{|n|\xi}}{m}\Phi(-|n|+1,m+1;\xi)\\
\Phi(-|n|,m;\xi) \end{array}\right),
\end{eqnarray}
where $m=|j|+1/2$, $\xi=r^2/2$, and
\begin{equation}
N_{n,j}^2=4\pi  \frac{ |n|! [(m-1)!]^2}{(m+|n|-1)!} .
\end{equation}
The matrix elements of $V^{\rm pert}$ in the basis of unperturbed Landau
levels, $\{ |n,j\rangle \}$,  are with $\xi_0=\frac12 (R/l_B)^2$ given by
\begin{equation} 
 \left\langle n,j|V^{\rm pert}|0,j\right\rangle= 
 -V_0\frac{4\sqrt2 \ \pi}{N_{0,j}N_{n,j}} \frac{\xi_0^m}{m} \Phi(|n|+m,m+1;-\xi_0).
\end{equation}

The integrated current to first order in $V_0$ then follows as
\begin{eqnarray}\nonumber
 I_{0,j<0} & = &v_F\int_0^{\infty}dr\;\Psi_{0,j}^{{\rm pert}\dagger}
\left(\begin{array}{cc}
0 & -ie^{-i\varphi} \\ ie^{i\varphi} & 0 \end{array}\right)
\Psi^{\rm pert}_{0,j} \\  
&=&\frac{v_F V_0}{l_B E_B} C_{0,j} + {\cal O}(V_0^2),
\end{eqnarray}
cf.~Eq.~\eqref{zerocurrent},
where $\Psi^{\rm pert}_{0,j}(r,\varphi)$ is determined by
Eqs.~\eqref{circstates} and \eqref{psipert}, and in the last line we have restored physical units.
The radial integral for the coefficient $C_{0,j}$ gives
\begin{equation}\label{cpert}
C_{0,j}(\xi_0)=  -\frac{\sqrt{2}}{\pi} \frac{\xi_0^m}{\Gamma(m+1)} 
\sum_{n=1}^\infty \frac{\Phi(n+m,m+1,-\xi_0)}{n},
\end{equation}
where we recall $m=|j|+1/2$.
This result is illustrated in Fig.~\ref{fig10} for two values of $R$ 
and features a peak at $j=j_0\approx-\xi_0$, see Eq.~\eqref{peakvalue}.


\begin{thebibliography}{99}
\bibitem{footnote}
Technically, the spinor solutions reported in Ref.~\cite{zhang}, see Eqs.~(5,6) therein, are each defined only up to an overall arbitrary factor. 
When those factors are properly taken into account, 
 instead of Eq.~(7) in Ref.~\cite{zhang}, one obtains our
  Eq.~\eqref{matching} which implicitly 
defines the dispersion relation.
\end{thebibliography}
\end{document}